\documentclass[sigconf,review,screen]{acmart}

\AtBeginDocument{%
  }

\acmConference[]{}{}

\usepackage[most]{tcolorbox}
\usepackage{multirow}
\usepackage{xspace}
\usepackage{enumitem}
\usepackage{pifont}

\newcommand{\tool}{\textsc{SliceMate}\xspace}
\newcommand{\bench}{\textsc{SliceBench}\xspace}

\setcopyright{none}

\begin{document}

\title{\tool: Accurate and Scalable Static Program Slicing via LLM-Powered Agents}

\author{
	\normalsize
	Jianming Chang\textsuperscript{1}, 
	Jieke Shi\textsuperscript{2}, 
	Yunbo Lyu\textsuperscript{2}, 
	Xin Zhou\textsuperscript{2}, 
	Lulu Wang\textsuperscript{1}, 
	Zhou Yang\textsuperscript{3}, 
	Bixin Li\textsuperscript{1}, 
	David Lo\textsuperscript{2}\\
	\textsuperscript{1}Southeast University, Nanjing, China 211189\\
	\textsuperscript{2}Singapore Management University, Singapore 188065\\
	\textsuperscript{3}University of Alberta, Edmonton AB T6G 2R3, Canada
}

\begin{abstract}
Static program slicing, which extracts the executable portions of a program that affect the values at a specific location, supports many software analysis tasks such as debugging and security auditing. However, traditional slicing tools rely on computationally expensive reachability analysis over dependency graphs, which struggle to scale to large programs and often fail to handle code with incomplete syntax. Recently emerged learning-based methods, while more robust to such cases, still fall short of achieving comparable performance to traditional methods on well-formed code.

In this work, we propose \tool, a novel static program slicing solution powered by Large Language Model (LLM) agents. It bypasses the need for explicit dependency graph construction and achieving superior slicing accuracy. Concretely, \tool integrates three specialized agents: (1)~a synthesis agent that produces candidate slices by incrementally expanding the scan scope across functions and files guided by LLM-inferred dependencies; (2)~a verification agent that performs conciseness and completeness checks of the candidate slices, detecting missing or irrelevant statements; and (3)~a refinement agent that repairs the slices with minimal edits in accordance with the verification results. These agents are orchestrated by a control module that ensures timely convergence and outputs high-quality slices without manual intervention. For rigorous evaluation, we construct a new and high-quality benchmark, \bench, comprising 2,200 manually annotated Java and Python programs, with program lengths ranging from 5 to 8,577 lines, significantly larger than those in existing slicing benchmarks. Experimental results show that \tool greatly outperforms both traditional and learning-based slicing tools. Compared to the best-performing baseline, \tool improves slicing accuracy by up to 22\% and F1 score by 28\%. Moreover, when applied to downstream debugging and bug localization tasks, \tool-generated slices boost Top-10 localization accuracy by up to 34\%.

\end{abstract}




\maketitle
\fancyhead{}  
\thispagestyle{empty}  
\section{Introduction}

Static program slicing~\cite{Weiser1981, Frank1995, xu2005brief} is a crucial analysis technique that identifies executable portions of code based on a user-specified slicing criterion (e.g., a variable or program point) without executing the program. It has proven effective in a wide spectrum of software engineering tasks, such as debugging~\cite{Ezekiel2021, binkley2014orbs, li2020more}, vulnerability detection~\cite{Cao2022, papotti2024effects, salimi2022vulslicer}, and patch validation~\cite{Vidziunas2024, al2025reduce, nong2024automated}, where isolating relevant code segments helps developers better understand program behavior and locate faults~\cite{Vidziunas2024, Xuan2021, chang2025bridging}.

\begin{figure}[!t]
	\centering
	\includegraphics[width=0.46\textwidth]{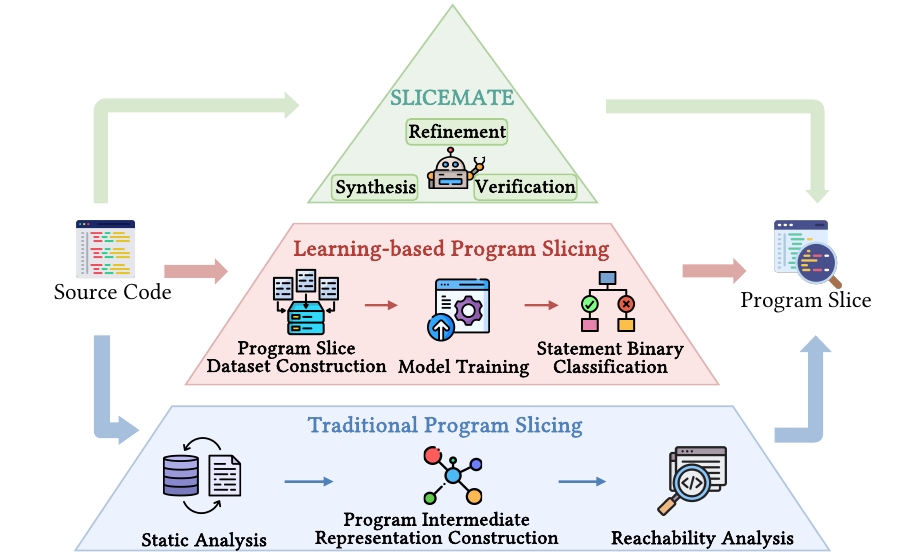}
	\vspace{-0.3cm}
	\caption{Difference between \tool and existing static slicing techniques. \tool produces program slices without requiring explicit dependency graph construction, large-scale training data, or task-specific fine-tuning.}
	\label{IntroExample}
	\vspace{-0.8cm}
\end{figure}

Static program slicing techniques can be broadly divided into two categories: traditional static slicing and learning-based slicing. Traditional methods (at the bottom of Figure~\ref{IntroExample}), which have been extensively studied in prior studies~\cite{Ferrante1987, Binkley2019, Weiser1979, Gallagher2025}, typically rely on reachability analysis over Program Dependency Graphs (PDGs)~\cite{Ferrante1987} extracted from the whole program. The slicing tools traverse the large dependency graphs from the slicing criterion and collect all reachable nodes that may influence the criterion's value, consuming substantial time and resources, which results in inherent scalability limitations in traditional methods~\cite{Gallagher2025}. Additionally, traditional slicing techniques are language-specific: they depend on custom parsers and analysis logic tailored to particular language syntax and semantics, for example, JavaSlicer~\cite{Galindo2022} and TyperSlicer~\cite{wang2020} work exclusively for Java. This tight coupling limits their applicability in modern software systems, where multiple languages are often used together. Moreover, these tools do not generalize well to incomplete or syntactically incorrect code, common during debugging or exploratory development, where dependency information may be missing or unreliable. Collectively, these limitations reduce the practicality of traditional slicing approaches in large-scale, multi-language development environments.

Recently, several studies~\cite{Yadavally2024, Yadavally2023, Kimya2024} have explored learning-based program slicing techniques by leveraging Large Language Models (LLMs), which have demonstrated strong performance across various programming tasks~\cite{hou2024large, wang2024software}, to address the limitations of traditional methods. As shown in Figure~\ref{IntroExample}, these approaches typically formulate program slicing as a statement-level binary classification task, where the model takes the slicing criterion along with each statement and predicts whether the statement should be included in the slice based on its relevance. Trained on large-scale datasets of programs and their slices, these models can infer dependencies without explicitly constructing dependency graphs, thereby mitigating scalability issues and enabling the handling of incomplete or syntactically incorrect code. Furthermore, by fine-tuning on language-specific datasets, they can generalize across diverse programming languages and codebases.

However, while learning-based methods mitigate some limitations of traditional slicing tools, they introduce challenges of their own. Notably, prior work~\cite{Yadavally2024, Yadavally2023} relies on compact models such as CodeBERT~\cite{feng2020codebert}, whose efficacy depends on large volumes of training data produced by traditional slicing tools. This reliance not only ties their effectiveness to training data quality, which may miss certain slicing scenarios, but also caps their performance at the level of the slicing tool for data generation. For instance, NeuralPDA~\cite{Yadavally2023} and NS-Slicer~\cite{Yadavally2024} are both trained on slices produced by JavaSlicer~\cite{Galindo2022}, inherently limiting their maximum achievable performance—they reach at most 85\% accuracy on benchmarks derived from JavaSlicer.
Furthermore, although these methods eliminate the need for dependency graph construction, they model slicing as a statement-level binary classification task, processing each statement individually. This approach becomes inefficient for large-scale programs, where models must process a corresponding number of statements line by line, still incurring high computational costs. Consequently, current learning-based methods do not fully resolve scalability bottlenecks, leaving room for optimization.

\textbf{Our Work}. We propose \tool, a novel static program slicing solution powered by LLM agents. As illustrated in Figure~\ref{IntroExample}, \tool fundamentally departs from traditional slicing tools by bypassing static analysis and dependency graph construction. It also overcomes key limitations of learning-based approaches, which rely on large-scale labeled data and frame slicing as a statement-level classification task. Instead, \tool reframes slicing as an LLM-driven code generation process that scales to large, multi-file programs and robustly handles incomplete or non-compilable code by leveraging the broad programming knowledge embedded in LLMs, without requiring task-specific training or fine-tuning.

Specifically, \tool takes program files and a slicing criterion as input, and generates the slice by orchestrating three specialized LLM agents:
(1)~a synthesis agent that generates candidate slices by incrementally expanding the scan scope across functions and files, guided by LLM-inferred inter-procedural dependencies, until the slice is deemed complete;
(2)~a verification agent that checks the completeness and conciseness of the candidate slices, identifying missing or irrelevant statements; and
(3)~a refinement agent that minimally edits the slices based on verification feedback to ensure correctness. Each agent operates under tailored prompts specifying its role, expected input and output formats, and task objectives, along with contextual information from the current slicing stage. \tool also incorporates a control module that coordinates the verification and refinement agents, driving convergence and producing high-quality slices without manual intervention.

To rigorously evaluate \tool, we manually construct a benchmark named \bench comprising 2,200 program slice instances from CodeNet and GitHub. Each instance includes the original code (ranging from 5 to 8,577 lines), a slicing criterion, and the ground-truth slice. To ensure the reliability of the slices, all annotations underwent dual independent review and agreement validation. Using \bench, we compare \tool against three popular traditional and two learning-based slicing techniques. Results show that \tool significantly outperforms all baselines—particularly on the large-scale programs with an average of 2,106 statements and multiple files—achieving up to 63.1\% higher accuracy and 62.5\% higher F1 score than the best-performing baseline. \tool requires only 0.15 US dollars per slice from GPT-4o API for large-scale programs. Furthermore, in downstream tasks such as program debugging and bug localization, slices generated by \tool improve the Top-5 and Top-10 localization accuracy of several existing bug localization tools by up to 32\% and 34\%, respectively, demonstrating strong real-world applicability.

We summarize our main contributions as follows:

\begin{itemize}[leftmargin=*]
	\item We present \tool, the first multi-agent framework  powered by LLMs for static program slicing, significantly improving scalability and generalizability.

	\item We introduce \bench, a new benchmark of 2,200 ground-truth slicing instances in Java and Python, covering both intra- and inter-procedural cases and large-scale programs. The dataset is publicly available to foster future research.

	\item We comprehensively evaluate \tool against multiple traditional and learning-based slicing tools, showing superior performance across various metrics and demonstrating its effectiveness in downstream tasks such as debugging and bug localization.
\end{itemize}
\section{Background}

\subsection{Static Program Slicing}

Slicing techniques are broadly classified into static and dynamic slicing~\cite{Kimya2024}, depending on whether runtime information is required. Our work focuses on static slicing, which identifies all statements that may affect a slicing criterion across possible execution paths without executing the program. Static slicing is more practical, especially when executing the program is difficult or infeasible. Moreover, based on the direction of analysis, static slicing can be further divided into backward and forward slicing~\cite{bergeretti1985information, reps1989interference}. Given a slicing criterion, backward slicing computes all statements that may influence it, while forward slicing identifies statements that may be affected by it. Backward slicing is widely used in tasks such as debugging and bug localization, where developers need to trace the source of a fault or understand how a variable is derived. In contrast, forward slicing is often applied in change impact analysis. Both slicing methods require analyzing intra-procedural dependencies, including data and control, as well as inter-procedural dependencies introduced by function calls~\cite{Horwitz1990}. Given the critical role of debugging and localization in software analysis, we specifically evaluate the effectiveness of backward slicing in these contexts.

\begin{figure*}[!ht]
	\centering
	\includegraphics[width=0.95\textwidth]{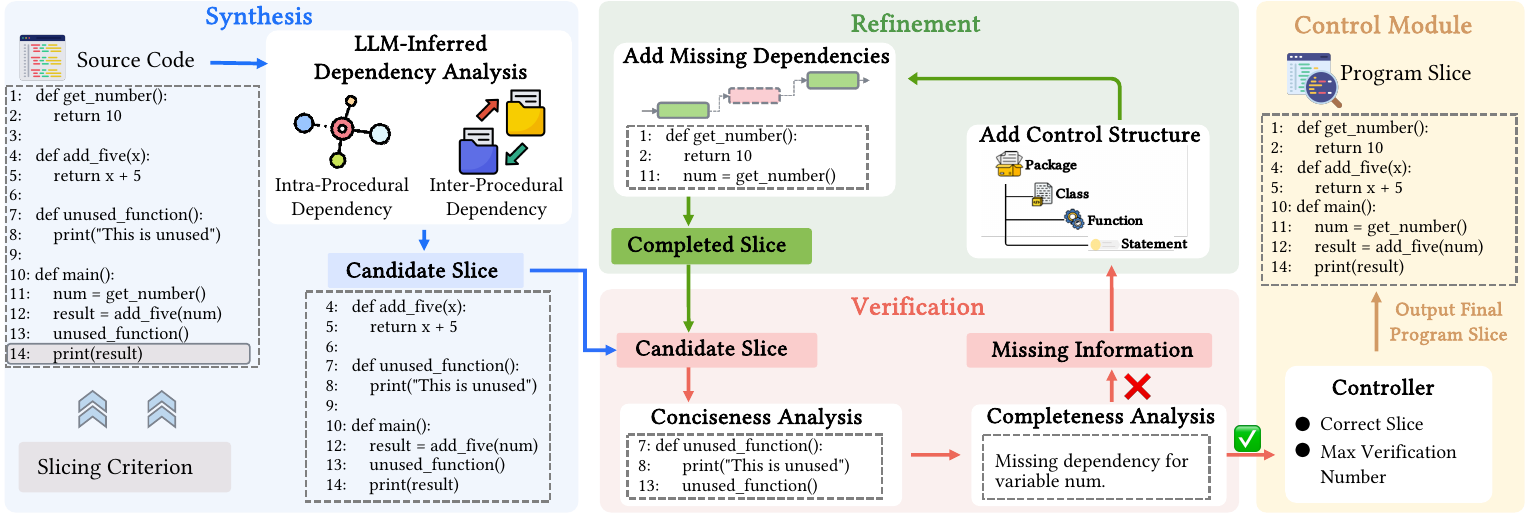}
	\vspace{-0.3cm}
	\caption{Overview of our approach. \tool generates a slice via synthesis agent and iteratively improves it through verification and refinement agents under control module.}
	\label{workflowPic}
	\vspace{-0.2cm}
\end{figure*}

\subsection{Agent for Software Engineering}

An agent is typically defined as an autonomous system that interacts with its environment to achieve specific goals through adaptive behavior~\cite{Wang2024, XI2025}. In software engineering, agents, especially those powered by LLMs, have been applied across various stages of the development lifecycle~\cite{Hou2024, He2025}, including requirement analysis~\cite{ren2024combining, hemmat2025research}, implementation~\cite{ma2025swe}, and testing~\cite{bouzenia2025you, tomic2025evaluation}. These agents autonomously coordinate workflows such as code analysis, tool invocation, and output refinement, delivering adaptive and context-aware solutions to real-world challenges in software engineering.

LLM-based software engineering agents typically follow one of two paradigms~\cite{Jin2025}. The first adopts fixed, domain-specific workflows, as exemplified by RepairAgent~\cite{Bouzenia2025} and AutoCodeRover~\cite{zhang2024autocoderover}, which delivers highly reliable performance but is often limited to narrow domains such as program repair. The second paradigm is open-ended, as seen in SWE-Agent~\cite{Yang2024} and CODEAGENT~\cite{Zhang2024}, where agents autonomously select actions and manage workflows for diverse tasks like code generation. While this flexibility allows broader adaptation to various programming scenarios, it often sacrifices predictability and reliability—especially in complex tasks that demand precise control over output quality and format~\cite{Jin2025, XI2025}.

\tool targets program slicing, a domain-specific task that demands high precision and output fidelity. To meet this need, \tool adopts a fixed-process architecture with clearly defined roles for slice synthesis, verification, and refinement. At the same time, it introduces bounded autonomy: agents dynamically expand the code scope and iteratively revise slices based on verification feedback. This hybrid design retains the robustness of structured workflows while offering the adaptability required to scale to multi-function, multi-file slicing tasks in real-world settings.

\vspace{-0.1cm}
\section{\tool Method}

We present \tool, a LLM-based multi-agent framework designed to statically generate accurate slices in real-world programs. Figure~\ref{workflowPic} illustrates the overview of our \tool. \tool incorporates three specialized agents: synthesis, verification, refinement, as well as a control module that orchestrates their interactions. We detail the roles and workflows of these agents below.

\vspace{-0.25cm}
\subsection{Synthesis Agent}

\begin{figure}[!t]
	\centering
	\includegraphics[width=0.47\textwidth]{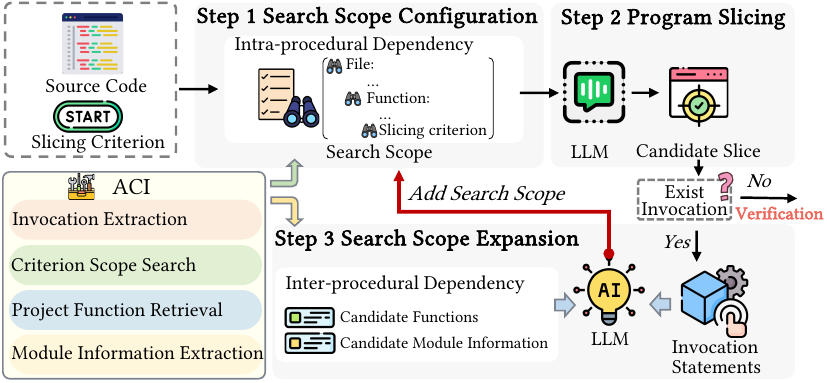}
	\vspace{-0.3cm}
	\caption{Workflow of the synthesis agent. It generates slices by incrementally expanding the analysis scope.}
    \label{generationModule}
\end{figure}

Figure~\ref{generationModule} illustrates the synthesis agent's workflow, which performs program slicing by iteratively expanding its search scope. Each iteration involves three steps: (1) configuring the current search scope, (2) generating a candidate slice, and (3) expanding the scope based on inferred dependencies. Although \tool does not need to construct explicit dependency graphs, it still requires information about code relationships to guide the slicing process. Therefore, to support intra-procedural and inter-procedural slicing, we design four Agent-Computer Interfaces (ACIs)~\cite{Yang2024} that help the agent retrieve relevant information as needed in each iteration:
\begin{itemize}[leftmargin=*]
	\item Criterion Scope Search. Determines the initial scope based on the slicing criterion. Using Tree-sitter\footnote{\url{https://tree-sitter.github.io}}, a robust parser generator tool, it returns the enclosing function body or surrounding code block depending on whether the criterion is a local variable or a global member like a class variable.
	\item Invocation Extraction. Parses the candidate slice to extract function calls by matching method invocation nodes in the syntax tree, excluding built-in functions (e.g., print), to help the agent locate corresponding definitions.
	\item Project Function Retrieval. Extracts all function declarations in the project along with enclosing class names, parameter lists, and inheritance context to understand available call targets.
	\item Module Information Extraction. Expands the context beyond the current function by returning referenced member/global variables, declarations, and the function bodies modifying them.
\end{itemize}

\begin{figure}[!t]
	\centering
	\includegraphics[width=0.48\textwidth]{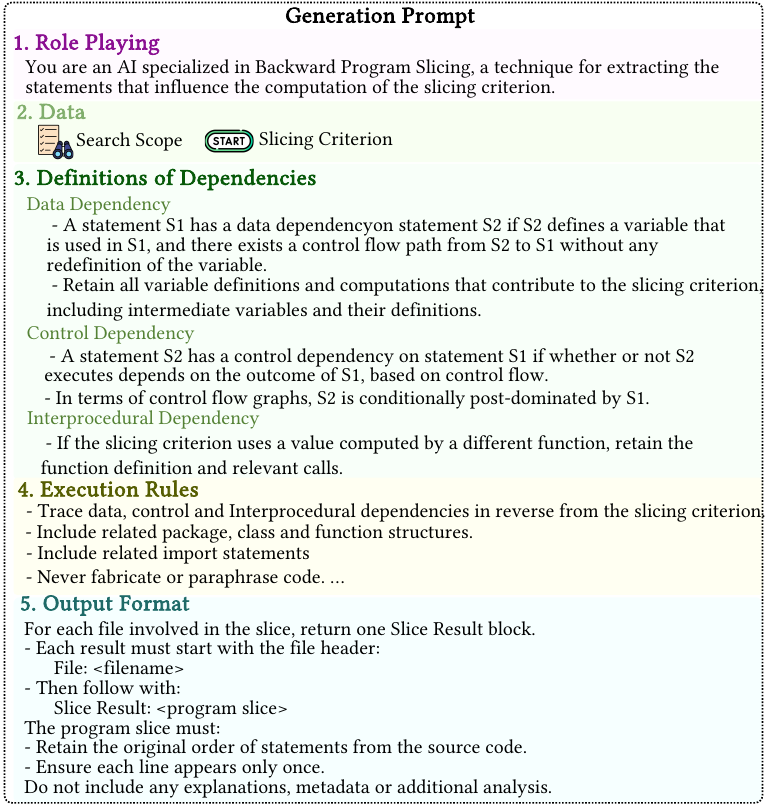}
	\vspace{-0.7cm}
	\caption{Prompt for the synthesis agent}
	\label{generationPrompt}
	\vspace{-0.3cm}
\end{figure}

With these ACIs, the synthesis agent can effectively gather the necessary information to perform inter-procedural slicing. The agent's workflow is as follows:

\vspace{0.15cm}
\noindent\textbf{Step 1: Search Scope Configuration.} Directly analyzing the entire source file is impractical, as these files are often too large to fit within the input constraints of LLMs and contain amounts of irrelevant code that may hinder effective reasoning. To overcome this, we adopt function-level granularity, a widely used scope in contextual code analysis \cite{Sajnani2016}, as the initial search boundary. Given a source program and slicing criterion information, including the corresponding file path and line number, the synthesis agent invokes the Criterion Scope Search interface. This interface identifies the slicing criterion based on the provided information and retrieves the code segment at the same syntactic structure level as the criterion to establish the initial search scope.

\vspace{0.15cm}
\noindent\textbf{Step 2: Program Slicing.} The agent leverages the LLM, guided by a structured prompt shown in Figure \ref{generationPrompt}. This prompt comprises five key components: (1) Role Assignment, which defines the LLM's role as a program slicing assistant responsible for identifying and extracting code segments relevant to the slicing criterion; (2) Input Data, specifying the model inputs, including the slicing criterion and its associated search scope; (3) Dependency Definitions, providing formal definitions and analytical methods for identifying program dependencies to enable accurate reasoning about code relationships; (4) Execution Rules, detailing how to traverse dependencies, the types of code elements to consider, and any special considerations; and (5) Output Format, prescribing a standardized structure for the output to ensure that the results include file names and their corresponding sliced code segments.

\begin{figure}[!t]
	\centering
	\includegraphics[width=0.48\textwidth]{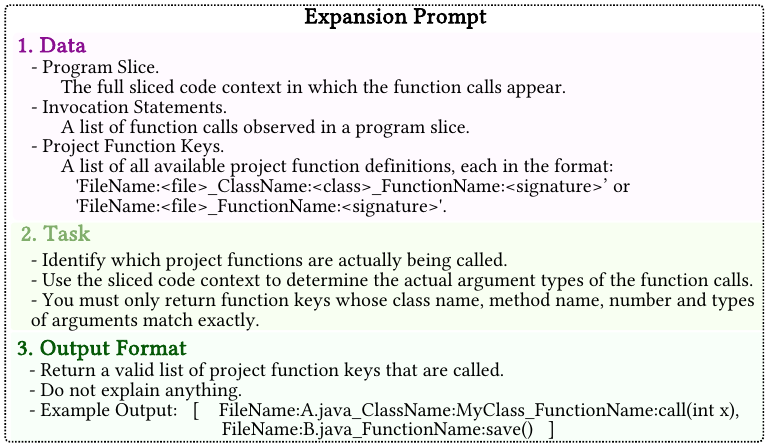}
	\vspace{-0.7cm}
	\caption{Prompt for the search scope expansion}
	\label{expandPrompt}
	\vspace{-0.45cm}
\end{figure}

\vspace{0.15cm}
\noindent\textbf{Step 3: Search Scope Expansion.} After generating the candidate slice, the synthesis agent expands the scope of dependency analysis.\ding{182} It first invokes the Invocation Extraction Interface to identify invocation statements within the candidate slice, which indicate external functions requiring further analysis. At the same time, the agent uses the Project Function Retrieval Interface to obtain comprehensive metadata about all functions in the project, including their associated classes, function names, and parameter lists. This metadata enables the agent to accurately distinguish complex features between functions like overloading or renaming. \ding{183} Based on the extracted invocation statements and the retrieved function list, the agent autonomously identifies relevant target functions using the prompt structure shown in Figure~\ref{expandPrompt}, and extracts the corresponding function bodies to expand the current search scope. To improve the LLM's understanding of the expanded context, the inserted code is annotated with the corresponding filenames. The enriched search scope is then used as input for a new round of program slicing. \ding{184} This iterative process continues until no additional invocations are found in the slicing results, indicating that the current function-level analysis scope is complete. To extend the slice beyond the function level, the agent invokes the Module Information Extraction Interface to retrieve code relevant to the current slice that is not contained within the current function, such as global variables, and structural elements like package and class declarations. Once the function-level expansion is finalized, this information is integrated into the analysis scope by filename, enabling the LLM to perform the final stage of dependency analysis. The finalized slice is then forwarded to the verification agent.

\vspace{-0.3cm}
\subsection{Verification Agent}

\begin{figure}[!t]
	\centering
	\includegraphics[width=0.48\textwidth]{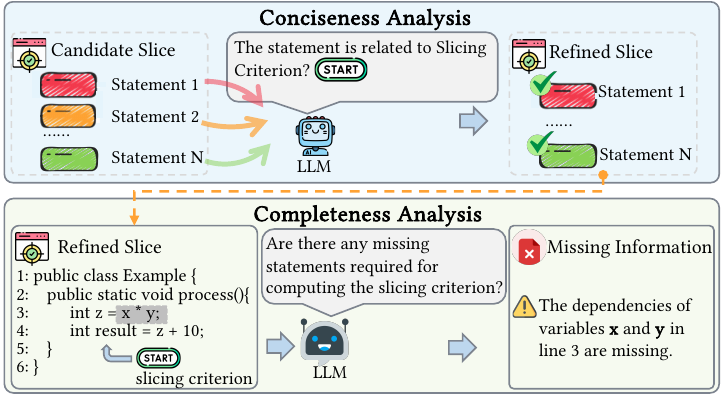}
	\vspace{-0.75cm}
	\caption{Workflow of the verification agent}
	\vspace{-0.5cm}
	\label{VerificationModule}
\end{figure}

Program slicing may produce incorrect results due to two main types of errors: missing relevant code (i.e., incompleteness) or including irrelevant code (i.e., lack of conciseness)~\cite{BINKLEY19961}. To mitigate both, we introduce a verification agent. As illustrated in Figure~\ref{VerificationModule}, this agent comprises two components: conciseness analysis and completeness analysis. Since static slices aim to capture the full set of reachable code, we perform completeness analysis after concise analysis to ensure that essential code segments are not inadvertently omitted from the final slice.

The conciseness analysis leverages the LLM's semantic understanding of code to assess whether each statement in the candidate slice is relevant to the computation of variables in the slicing criterion. The agent also filters out unnecessary statements, such as \texttt{print} commands. The prompt designed for this task consists of three components: (1) Role Playing, which assigns a specific role to the LLM—serving as a program analyst responsible for removing redundant lines of code from the slice; (2) Data, which provides the model with the candidate slice and the slicing criterion; and (3) Task, which instructs the model to analyze dependencies and determine whether each line contributes to the slicing criterion. Additionally, since static slicing typically operates at the line level, we explicitly instruct the model not to remove partial line content, in order to preserve the completeness of each line. The output of the LLM is a refined slice with redundant statements removed.

\begin{figure}[!t]
	\centering
	\includegraphics[width=0.48\textwidth]{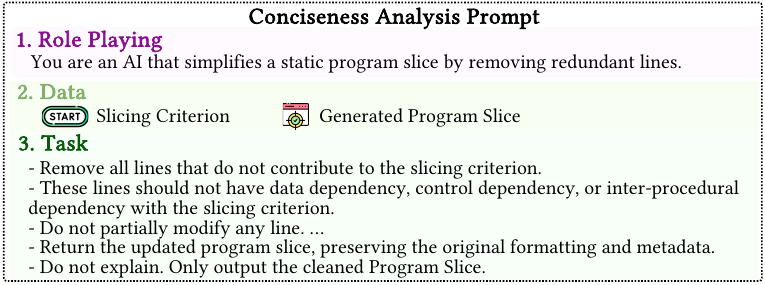}
	\vspace{-0.7cm}
	\caption{Prompt for the conciseness analysis}
	\label{VerRendundancyPrompt}
	\vspace{-0.5cm}
\end{figure}

Completeness analysis is to determine whether any necessary code has been omitted that could affect the accuracy of the current slice. The model conducts this analysis by comparing the candidate slice against the searched scope to ensure that all required dependency paths leading to the slicing criterion are preserved. The prompt used for this purpose consists of four components: (1) Role Playing, which assigns the LLM the role of an inspector responsible for verifying the completeness of the slice; (2) Data, which supplies the necessary inputs to the model, including the slicing criterion, the searched scope, and the candidate slice; (3) Task, which instructs the model to identify missing dependencies, while also verifying the structural integrity of the slice to ensure that critical contextual elements—such as class or function definitions—are retained; and (4) Output Format, which defines the format of the model's response, enabling it to return a list of missing dependencies and structural elements required for completing the slice.

\begin{figure}[!t]
	\centering
	\includegraphics[width=0.48\textwidth]{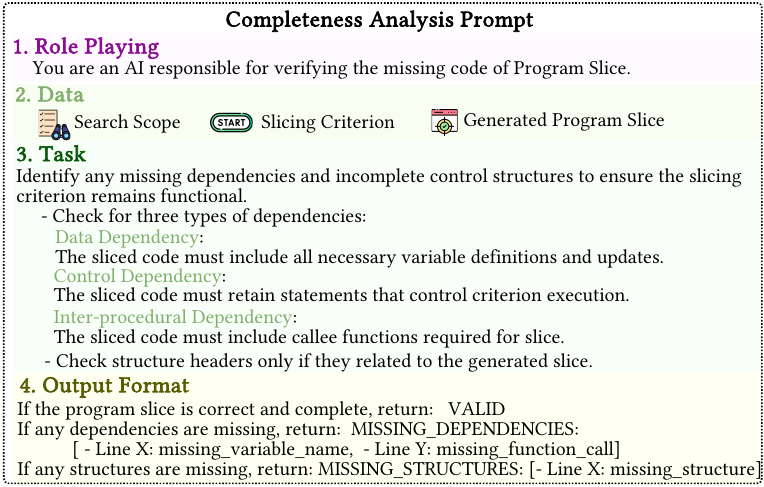}
	\vspace{-0.7cm}
	\caption{Prompt for the completeness analysis}
	\label{VerCorrectnessPrompt}
	\vspace{-0.83cm}
\end{figure}

\subsection{Refinement Agent}

\begin{figure}[!t]
	\centering
	\includegraphics[width=0.48\textwidth]{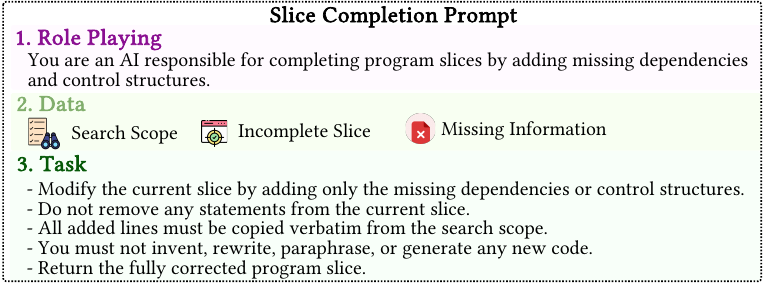}
	\vspace{-0.7cm}
	\caption{Prompt for the refinement agent}
	\label{compeletionPrompt}
\end{figure}

Since the model is more adept at processing code with an individual role \cite{Jin2025}, we decouple the slice completion task into an independent agent. This agent is responsible for reconstructing the corresponding slice based on the missing information identified during verification. The designed prompt consists of three components, as shown in Figure \ref{compeletionPrompt}: (1) Role Playing, which assigns the LLM the role of slice completer; (2) Data, which provides the analysis context, including the candidate slice, the searched scope, and the identified missing elements; and (3) Task, which instructs the model to complete the slice strictly based on the missing information. To prevent the LLM from generating arbitrary or invented code, we explicitly constrain it to reuse only content from the original code within the searched scope.

\vspace{-0.2cm}
\subsection{Control Module}

The control module manages the verification and refinement process to prevent infinite loops. It terminates the slicing workflow if the verification result shows the slice is already valid or if the number of verification-refinement iterations exceeds a predefined threshold. Based on empirical results (see Section~\ref{subsec:Control}), we set the maximum number of iterations to 5, which provides a reliable safeguard while ensuring convergence.

\vspace{-0.1cm}
\section{\bench Construction}
\label{sec:dataset}

While program slicing has been extensively studied and several benchmarks have been proposed~\cite{Ahmed2021, Binkley2014, Galindo2022, Yadavally2024}, most focus on dynamic slicing, and the available static slicing benchmarks~\cite{Galindo2022, Yadavally2024} are typically constructed by aggregating outputs from traditional slicing tools without sufficient human verification. As a result, these benchmarks cannot guarantee slice correctness and often miss critical elements such as import statements and cross-file dependencies. To address this limitation, we manually construct \bench with high-quality static slicing annotations.
\vspace{-0.2cm}

\subsection{Program Selection}
To ensure consistency with the existing benchmark \cite{Yadavally2024}, we follow the same data source and extract code from CodeNet dataset \cite{puri2021}, which includes over 4,000 programming problems and 14 million corresponding solutions. Given the high cost of manual annotation and the infeasibility of processing the entire dataset, we adopt a round-robin sampling strategy to select 1,000 representative programs. Specifically, we iteratively sample the first solution for each problem. After all problems have one solution selected, we proceed to the second solution for each, and so on, until reaching 1,000 programs to serve as slicing targets. For each program, we designate the final print statement as the slicing criterion, as it typically encapsulates the program’s overall logic and output behavior.

To evaluate the cross-language generalizability of \tool, we further sample 1,000 Python and 1,000 Java programs from CodeNet, yielding 2,000 slicing instances. However, CodeNet programs are relatively short, averaging only 30 lines, and usually consist of a single function. To address this limitation, we additionally collect real-world programs from open-source GitHub repositories. For Java, we use the Defects4J dataset \cite{Just2014}; for Python, the BugsInPy dataset \cite{Widyasari2020}. Both are widely used bug benchmarks comprising real-world buggy programs and their associated failing test units. For each language, we select the top 100 instances, averaging 2,106 code statements. In these cases, we use the assert statement in the failing test unit as the slicing criterion and treat the corresponding code files as slicing targets.
\vspace{-0.2cm}

\subsection{Manual Construction of Program Slice}
The program slice selection process is conducted by a team of eight experienced annotators, including two professors with research expertise in program slicing, two PhD candidates, and four Master's students. Each annotator has at least four years of programming experience and a solid understanding of program slicing theory.

\vspace{0.1cm}
\noindent\textbf{Annotation Guidelines.} Following standard static slicing principles, annotators are instructed to select statements from the source code that contribute to the slicing criterion through data dependencies, control dependencies, or inter-procedural dependencies. Both direct and indirect dependencies are considered during annotation.

\vspace{0.1cm}
\noindent\textbf{Annotation Procedure.} Given a source program and a specific slicing criterion, each annotator first identifies the variables involved in the criterion. Then, guided by the three types of dependencies, they select relevant code statements that influence the computation of those variables. After completing their selections, we compute the Cohen's Kappa score~\cite{Rastkar2010} between the two annotators to assess agreement. For the CodeNet programs, the average Kappa score is 0.927, indicating almost perfect agreement. For the GitHub programs, the average score is 0.893, demonstrating high consistency even in more complex, real-world codebases. Although inter-annotator agreement is high, some disagreements remain. In such cases, one of the professors reviews both slices and makes the final decision on the correct slice.

\vspace{-0.3cm}

\section{Experiment Setup}
Our evaluation considers the following research questions:

\vspace{-0.3cm}
\subsection{Research Questions}
\begin{itemize}[leftmargin=*]
	\item \textbf{RQ1: Is \tool more accurate than existing program slicing tools?} We evaluate the slicing performance of \tool by comparing it with state-of-the-art static slicing techniques on the \bench benchmark.
	\item \textbf{RQ2: How effective are \tool-generated slices in downstream applications?} We assess how well slices produced by \tool support practical tasks such as program debugging and bug localization, compared to slices from existing tools.

    \item \textbf{RQ3: What are the contributions of \tool's key components?}
We perform ablation studies to examine the impact of each component on the overall performance of \tool.
\end{itemize}

\begin{table*}[t!]
	\centering
	\caption{\tool vs. other program slicing tools on \bench (RQ1).}
	\vspace{-0.4cm}
	\small
	\setlength{\tabcolsep}{6pt}
	\begin{tabular}{lccccc ccccc}
		\hline \hline
		\multirow{2}{*}{Tool} &
		\multicolumn{5}{c}{\bench-CodeNet} &
		\multicolumn{5}{c}{\bench-GitHub} \\ \cline{2-11}
		& Prec. & Rec. & F1 & Acc-EM & Acc. & Prec. & Rec. & F1 & Acc-EM & Acc. \\
		\hline
		\multicolumn{11}{c}{Python} \\ \hline
		Joern          & 0.950 & 0.688 & 0.760 & 0.206 & 0.707 & 0.199 & 0.193 & 0.136 & 0     & 0.492 \\
		NS-Slicer      & 0.892 & 0.554 & 0.661 & 0.067 & 0.575 & 0.186 & 0.520 & 0.222 & 0     & 0.558 \\
		NS-Slicer Pro  & 0.921 & 0.936 & 0.917 & 0.479 & 0.890 & 0.172 & \textbf{0.793} & 0.247 & 0     & 0.312 \\
		\tool   & \textbf{0.975} & \textbf{0.952} & \textbf{0.955} & \textbf{0.693} & \textbf{0.945} & \textbf{0.931} & 0.689 & \textbf{0.761} & \textbf{0.020} & \textbf{0.943} \\
		\hline
		\multicolumn{11}{c}{Java} \\ \hline
		Joern          & 0.948 & 0.656 & 0.752 & 0.006 & 0.707 & 0.296 & 0.226 & 0.187 & 0     & 0.490 \\
		Javaslicer     & 0.884 & 0.827 & 0.835 & 0.304 & 0.826 & 0.644 & 0.367 & 0.392 & 0     & 0.541 \\
		TypeSlicer     & 0.932 & 0.801 & 0.842 & 0.003 & 0.812 & 0.734 & 0.398 & 0.447 & 0     & 0.764 \\
		NS-Slicer      & 0.837 & 0.520 & 0.621 & 0.014 & 0.608 & 0.338 & 0.356 & 0.262 & 0     & 0.590 \\
		NS-Slicer Pro  & 0.894 & 0.924 & 0.895 & 0.409 & 0.885 & 0.319 & \textbf{0.917} & 0.408 & 0     & 0.390 \\
		\tool   & \textbf{0.948} & \textbf{0.944} & \textbf{0.936} & \textbf{0.541} & \textbf{0.922} & \textbf{0.896} & 0.675 & \textbf{0.736} & \textbf{0.060} & \textbf{0.895} \\
		\hline\hline
	\end{tabular}
	\label{Exp1ResultsTable}
	\vspace{-0.4cm}
\end{table*}

\vspace{-0.3cm}
\subsection{Evaluation Dataset}

We use our benchmark \bench, introduced in Section~\ref{sec:dataset}, to evaluate the accuracy of program slicing tools. For RQ2, which assesses the effectiveness of generated slices in downstream applications, we adopt the widely used Defects4J dataset for Java and BugsInPy for Python in the context of debugging and bug localization tasks. Both datasets include bug descriptions and are annotated with the corresponding buggy statements.

\vspace{-0.1cm}
\subsection{Baselines}
We compare \tool against the five static program slicing baselines, three traditional slicing tools and two learning-based slicing tools, as follows:
\begin{itemize}[leftmargin=*]
    \item Joern \cite{Yamaguchi2014} is a representative static slicing tool that performs dependency reachability analysis by constructing a code property graph. It supports both Java and Python.
    \item JavaSlicer \cite{Galindo2022} is a state-of-the-art program slicing tool for Java. It constructs a system dependency graph and performs reachability analysis to generate program slices.
    \item TyperSlicer \cite{wang2020} is another representative program slicing tool for Java. It constructs a sub-statement-level dependency graph that incorporates variable type analysis for identifying the slices.
    \item NS-Slicer \cite{Yadavally2024} is the state-of-the-art learning-based program slicing tool for Java and Python languages. It trains GraphCodeBERT \cite{Guo2021} based on JavaSlicer output data and performs program slicing by binary classification of each code statement.
    \item NS-Slicer Pro is an enhanced version of NS-Slicer that incorporates additional training data from our \bench.
\end{itemize}

\vspace{-0.1cm}
\subsection{Metrics}
Following prior work~\cite{Yadavally2024}, we evaluate the accuracy of program slices using the following metrics: \textit{Accuracy} = $\frac{\text{TP} + \text{TN}}{\text{TP} + \text{FP} + \text{FN} + \text{TN}}$, \textit{Precision} = $\frac{\text{TP}}{\text{TP} + \text{FP}}$, \textit{Recall} = $\frac{\text{TP}}{\text{TP} + \text{FN}}$, \textit{F1 score} = $\frac{2 \cdot \text{Precision} \cdot \text{Recall}}{\text{Precision} + \text{Recall}}$, and Exact-Match Accuracy (Accuracy-EM), which counts the number of cases where the generated slice exactly matches the ground-truth slice. All metrics are computed at the statement level. To reduce the impact of formatting variations, we exclude non-semantic symbols such as parentheses and colons from the evaluation.

For slice-based program debugging, we adopt the Ratio metric, also following~\cite{Yadavally2024}. Specifically, $Ratio_{1}$ measures the proportion of bugs whose corresponding slices contain at least one buggy statement, while $Ratio_{All}$ measures the proportion of bugs whose slices include all buggy statements. These metrics reflect the effectiveness of slicing tools in isolating bug-relevant code.

For bug localization, we use the standard Top-N metric~\cite{Yang2024ICSE}, which measures the number of bugs for which at least one buggy element appears within the top-N ranked positions (N = 1, 3, 5, 10).
\section{Experiment Results}

\subsection{RQ1. Overall Performance of \tool}
\subsubsection{Settings} We evaluate the accuracy of \tool against four baselines on the \bench dataset. Since NS-Slicer is fine-tuned using JavaSlicer outputs, we continue fine-tuning it on the \bench  for a fair comparison, referred to as NS-Slicer Pro. To mitigate the influence of randomness in NS-Slicer Pro's performance, we adopt a 10-fold cross-validation strategy, partitioning \bench into 10 folds. For each fold, we follow the original NS-Slicer training setup by splitting the data into training, validation, and test sets in an 8:1:1 ratio and report the average of the 10 test results. Additionally, \tool employs GPT-4o, the best model for program slicing \cite{Kimya2024}, with the temperature parameter set to 0 to minimize randomness in slice generation.

\vspace{-0.1cm}
\subsubsection{Results} The detailed results of the five program slicing tools are presented in Table \ref{Exp1ResultsTable}, which reports Precision, Recall, F1 score, Accuracy, and Accuracy-EM for each tool. The bolded value in the table indicate the highest value in each column. The results demonstrate that \tool consistently outperforms all baselines in both Java and Python, effectively handling CodeNet programs and real-world cross-file code from GitHub.

For the CodeNet dataset with relatively small code sizes, \tool achieved the highest scores across all evaluation metrics. Specifically, for Python code, the scores reached 0.975 (Precision), 0.952 (Recall), 0.955 (F1), 0.945 (Accuracy), and 0.693 (Accuracy-EM), while for Java code, the corresponding values were 0.948, 0.944, 0.936, 0.922, and 0.541, respectively. Compared to the best-performing baseline, NS-Slicer Pro, \tool improved the F1 score, Accuracy, and Accuracy-EM by 3.8\%, 5.5\%, and 21.4\% on Python code, and by 4.1\%, 3.7\%, and 13.2\% on Java code. Notably, on the larger-scale GitHub dataset, where programs are significantly more complex, \tool exhibited even greater performance gains. It consistently outperformed other tools in Precision, F1 score, Accuracy, and Accuracy-EM; although its Recall was lower than that of NS-Slicer Pro, \tool still achieved outstanding overall performance. For Python code from GitHub, it surpassed the best baseline by 51.4\% in F1 score, 38.5\% in Precision, and 2\% in EM Accuracy; for Java code, the improvements were 28.9\%, 13.1\%, and 6\%, respectively, in the same metrics.

\begin{figure*}[!t]
	\centering
	\includegraphics[width=0.99\textwidth]{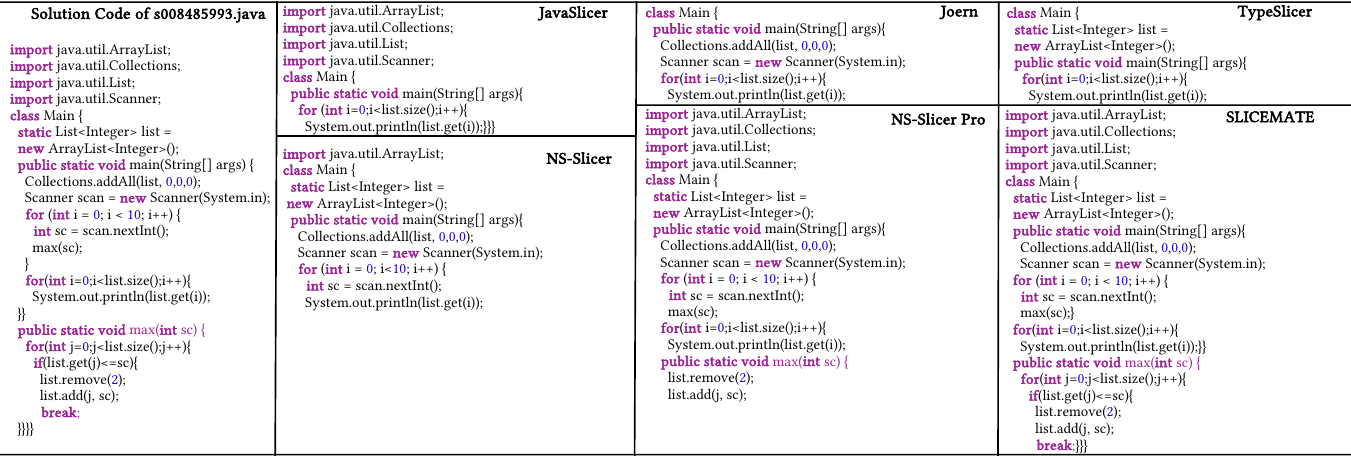}
	\vspace{-0.1cm}
	\caption{An illustrative example to show how \tool generates a slice compared with the baselines (RQ1).}
	\label{Exp1IntraSlice}
	\vspace{-0.3cm}
\end{figure*}

\vspace{-0.2cm}
\subsubsection{Analysis} Notably, \tool's advantages become more pronounced on large-scale programs. In small-scale cases, where all code lies within a single function or procedure, most tools have access to the complete dependency scope and thus perform similarly—though \tool still leads due to its finer-grained dependency reasoning. In contrast, large-scale programs often span multiple functions and files, where inter-procedural analysis is essential. For example, Figure~\ref{Exp1IntraSlice} presents a CodeNet solution involving a static \texttt{list}, a \texttt{main} function, and a \texttt{max} function that updates the \texttt{list}. When slicing based on a print statement in \texttt{main}, traditional tools fail to include \texttt{max} due to limited inter-procedural reachability. While NS-Slicer variants include \texttt{max}, their coarse-grained models miss critical statements. In contrast, \tool captures the data dependency via the updated \texttt{list} and returns a complete, accurate slice. Additionally, it handles broader contexts like \texttt{import} statements and prunes irrelevant control branches—e.g., discarding an unused \texttt{else} branch—further boosting its Accuracy-EM.

On large-scale code, \tool exhibits strong performance in analyzing dependencies across multiple files, leading to higher overall accuracy. Although NS-Slicer Pro achieves the highest recall, its precision is significantly lower due to its statement-level binary classification strategy, which struggles with long-range dependencies and often includes irrelevant statements. In cross-file slicing scenarios, precisely matching the ground truth is inherently challenging, resulting in low Accuracy-EM scores across all tools. Nevertheless, \tool consistently achieves high F1 and accuracy scores, owing to its function-aware design that facilitates more precise contextual understanding. These results indicate that \tool produces slices that are both comprehensive and accurate.

\vspace{-0.1cm}
\begin{tcolorbox}[
	colback=gray!10, 
	colframe=black!70, 
	coltitle=black, 
	boxrule=0.75pt, 
	rounded corners, 
	drop shadow, 
	enhanced, 
	shadow={1mm}{-1mm}{0mm}{black!50}, 
	boxsep=0.1mm 
	]
	\textbf{Answer to RQ1:} \tool outperforms all slicing baselines in generating accurate program slices. On small-scale programs, it improves F1, Accuracy, and Accuracy-EM by up to 31.5\%, 37\%, and 62.6\%, respectively. On large-scale programs, the improvements are even more evident, with gains of up to 62.5\% in F1, 63.1\% in Accuracy, and 6\% in Accuracy-EM.
\end{tcolorbox}
\vspace{-0.3cm}

\subsection{RQ 2. Applications of \tool}

\begin{figure*}[!t]
	\centering
	\includegraphics[width=0.95\textwidth]{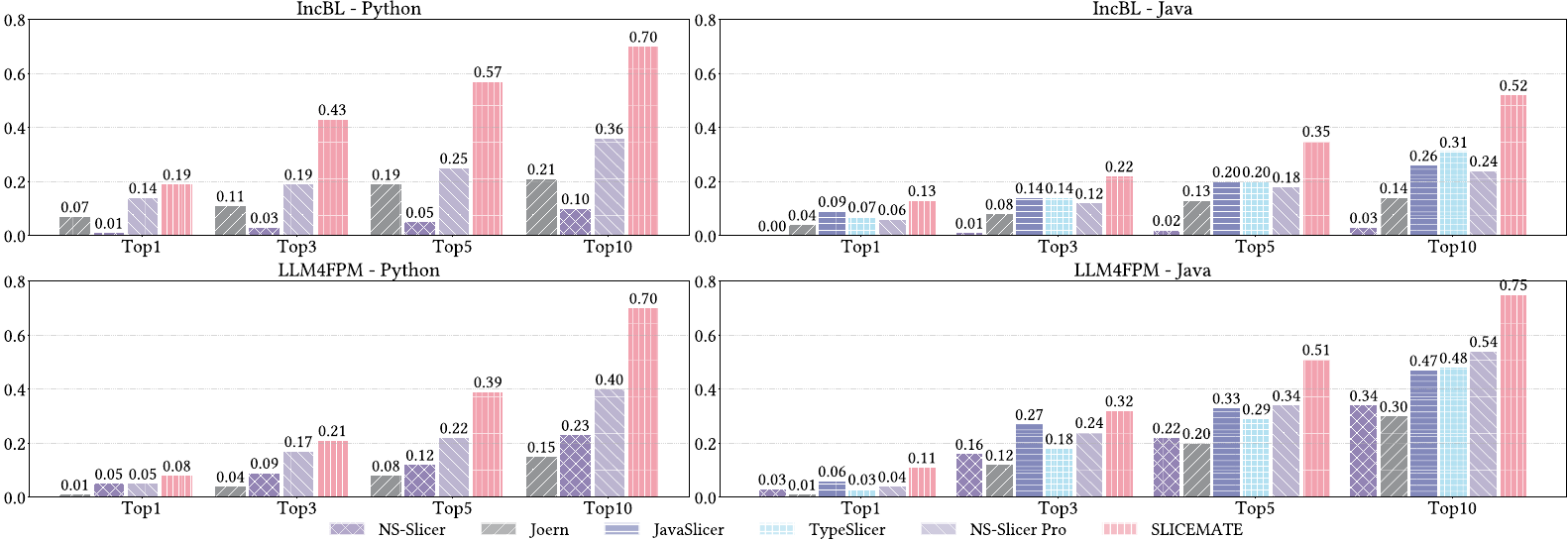}
	\vspace{-0.1cm}
	\caption{The bug localization performance of IncBL and LLM4FPM with \tool and other slicing baselines (RQ2). }
	\label{LocalizationResult}
	\vspace{-0.3cm}
\end{figure*}

\subsubsection{Settings} This experiment evaluates the practical effectiveness of \tool across representative applications. Program debugging is one of the earliest and most fundamental use cases of program slicing. Following prior experimental setups\cite{Yadavally2024}, we use the Ratio metric to assess effectiveness by determining whether the slice contains the buggy code. Additionally, considering that bug localization is a popular application of program slicing, we select two representative tools as application scenarios: IncBL \cite{Yang2022}, a classical bug localization method based on text similarity, and LLM4FPM \cite{Chen2025}, the latest approach that leverages LLM and program slicing to find the bug. The experiment leverages the slices generated by different program slicing tools to support the two bug localization methods. The effectiveness of each slicing tool in bug localization is evaluated using the Top-N.

\begin{table}[!t]
	\centering
	\caption{The debugging performance of \tool and the baselines (RQ2).}
	\vspace{-0.3cm}
	\small
	\begin{tabular}{llcc}
		\hline\hline
		Language & Tool & Ratio$_\text{All}$ & Ratio$_1$ \\
		\hline
		\multirow{4}{*}{Python}
		& Joern & 0.11 & 0.21 \\
		& NS-Slicer & 0.13 & 0.44 \\
		& NS-Slicer Pro & 0.60 & 0.79 \\
		& \tool & \textbf{0.71} & \textbf{0.91} \\
		\hline
		\multirow{6}{*}{Java}
		& JavaSlicer & 0.70 & 0.76 \\
		& Joern & 0.35 & 0.43 \\
		& TypeSlicer & 0.53 & 0.58 \\
		& NS-Slicer & 0.36 & 0.62 \\
		& NS-Slicer Pro & 0.83 & 0.92 \\
		& \tool & \textbf{0.84} & \textbf{0.93} \\
		\hline\hline
	\end{tabular}
	\label{ProgramDebuggingResults}
    \vspace{-0.6cm}
\end{table}

\vspace{-0.1cm}
\subsubsection{Results} Table \ref{ProgramDebuggingResults} presents the performance of each slicing tool in program debugging. The upper portion of the table reports the results on the BugsInPy dataset for Python, while the lower portion shows the results on the Defects4J dataset for Java. Bold values indicate the best performance in each column. According to the experimental results, \tool achieves the highest performance across both datasets, with a Ratio$_1$ of 0.91 and Ratio$_{All}$ of 0.71 on BugsInPY, and a Ratio$_1$ of 0.93 and Ratio$_{All}$ of 0.84 on Defects4J.

\begin{table*}[htbp]
	\centering
	\caption{Ablation study of verification and refinement modules (RQ3). $\Delta$ denotes the improvement of SliceMate over SliceMate-.}
	\vspace{-0.3cm}
	\small
	\setlength{\tabcolsep}{6pt}
	\begin{tabular}{cccccc ccccc}
		\hline \hline
		\multirow{2}{*}{Tool} &
		\multicolumn{5}{c}{\bench-CodeNet} &
		\multicolumn{5}{c}{\bench-GitHub} \\ \cline{2-11}
		& Prec. & Rec. & F1 & Acc-EM & Acc. & Prec. & Rec. & F1 & Acc-EM & Acc. \\
		\hline
		\multicolumn{11}{c}{Python} \\ \hline
		\tool-   & 0.921 & 0.902 & 0.900 & 0.246 & 0.883 & 0.761 & 0.398 & 0.470 & 0     & 0.888 \\
		\tool    & 0.975 & 0.952 & 0.955 & 0.693 & 0.945 & 0.931 & 0.689 & 0.761 & 0.020 & 0.943 \\
		$\Delta$       & +0.054 & +0.050 & +0.055 & +0.447 & +0.062 & +0.170 & +0.291 & +0.291 & +0.020 & +0.055 \\
		\hline
		\multicolumn{11}{c}{Java} \\ \hline
		\tool-   & 0.919 & 0.857 & 0.875 & 0.005 & 0.842 & 0.751 & 0.477 & 0.529 & 0     & 0.831 \\
		\tool    & 0.948 & 0.944 & 0.936 & 0.541 & 0.922 & 0.896 & 0.675 & 0.736 & 0.060 & 0.895 \\
		$\Delta$         & +0.029 & +0.087 & +0.061 & +0.536 & +0.080 & +0.145 & +0.198 & +0.207 & +0.060 & +0.064 \\
		\hline \hline
	\end{tabular}
	\label{ablationTable}
	\vspace{-0.3cm}
\end{table*}

Figure \ref{LocalizationResult} illustrates the effectiveness of different program slicing tools in bug localization. The x-axis represents the Top-N (N=1, 3, 5, 10), and the y-axis shows the value of Top-N. Among the evaluated tools, both IncBL and LLM4FPM, when supported by \tool, consistently achieve the best results on BugsInPY and Defects4J. Specifically, on BugsInPY, IncBL with \tool attains Top-N (N=1, 3, 5, 10) scores of 0.19, 0.43, 0.57, and 0.70, while LLM4FPM with \tool achieves 0.08, 0.21, 0.39, and 0.70. On Defects4J, IncBL reaches Top-N values of 0.13, 0.22, 0.35, and 0.52, and LLM4FPM achieves 0.11, 0.32, 0.51, and 0.75. Compared to the best-performing baselines, \tool improves IncBL's Top-N (N=1, 3, 5, 10) accuracy on BugsInPY by 5\%, 24\%, 32\%, and 34\%, respectively, and enhances LLM4FPM's performance by 3\%, 4\%, 17\%, and 30\%. On Defects4J, the improvements are 4\%, 8\%, 15\%, and 21\% for IncBL, and 5\%, 5\%, 17\%, and 21\% for LLM4FPM, respectively.

\vspace{-0.1cm}
\subsubsection{Analysis}
\textbf{Program Debugging.}
To investigate the effectiveness of traditional slicing tools, we analyze the successful cases in which they correctly identify the buggy statements. In these instances, the original programs average 1,878 lines of code, while the failed source code averages 2,620 lines, indicating that the effectiveness of traditional slicing tools declines in more complex codebases. In addition, slices produced by the traditional tools cover an average of 4 functions, compared to 6 functions in both the slices generated by \tool and the ground-truth slices. This discrepancy suggests the traditional tool may fail to capture the inter-procedural dependencies, contributing to their relatively lower Ratio scores. Furthermore, although NS-Slicer Pro achieved performance comparable to that of \tool, our analysis shows that its slices contained an average of 1,034 code statements, while the \bench dataset shows that the essential scope in Defects4J and BugsInPy averages around 112 statements. This indicates that NS-Slicer Pro tends to generate overly large slices, incorporating many irrelevant statements alongside the critical buggy code.

\vspace{0.1cm}
\noindent\textbf{Bug Localization.}
The program slices generated by \tool are notably more effective in assisting bug localization tools. Compared with other slicing tools, \tool achieves higher coverage of buggy statements, offering more opportunities for successful localization. While tools such as NS-Slicer Pro and JavaSlicer also attain relatively high coverage, their effectiveness is not guaranteed. For instance, NS-Slicer Pro fails to yield strong localization performance—sometimes even underperforming traditional tools like TypeSlicer. This is likely due to that its average slice size reaches 1,034 lines, compared to just 93 and 54 for \tool and TypeSlicer, respectively. The large volume of irrelevant code in NS-Slicer Pro's slices disrupts textual similarity-based localization techniques.

In contrast, \tool generates slices that are both precise and concise, preserving key buggy code while filtering out unrelated content. This makes the slices more suitable for localization tools that rely on contextual understanding of code. To further evaluate \tool’s impact, we compare the unique bugs identified in the Top-5 localization results~\cite{Yang2024ICSE} using \tool versus baseline slicing methods. We find that IncBL with \tool uniquely identifies 56 bugs in BugsInPy and 14 in Defects4J, while LLM4FPM with \tool uncovers 42 and 26 unique bugs, respectively, that are not detected with any of the baselines. These results highlight \tool’s practical effectiveness in improving bug localization.

\vspace{-0.1cm}
\begin{tcolorbox}[
	colback=gray!10, 
	colframe=black!70, 
	coltitle=black, 
	boxrule=0.75pt, 
	rounded corners, 
	drop shadow, 
	enhanced, 
	shadow={1mm}{-1mm}{0mm}{black!50}, 
	boxsep=0.1mm 
	]
	\textbf{Answer to RQ 2:} \tool demonstrates strong performance in supporting program debugging and bug localization. In debugging, it improves upon the best baseline by up to 12\% in bug coverage. For localization, integrating \tool with IncBL and LLM4FPM yields consistent improvements across all Top-N settings, reaching up to 34\%.
\end{tcolorbox}
\vspace{-0.3cm}

\subsection{RQ 3. Ablation Study}

\subsubsection{Settings} This experiment examines the impact of verification and refinement agents on \tool. Since these agents iteratively refine slices with complementary roles, they are removed together to assess their effect on accuracy. To determine the optimal number of iterations for the Control module, the experiment tests iteration counts from 1 to 10 and selects the one that maximizes the F1 score, a comprehensive performance measure. Given the need for more extensive verification and refinement in large-scale programs, we conduct the experiment on \bench-GitHub.

\vspace{-0.1cm}
\subsubsection{Effect of Verification and Refinement Agents} Table \ref{ablationTable} compares the performance of \tool-, which excludes the verification and refinement agents, with the full \tool, which includes both agents. The results show that integrating these agents consistently improves all evaluation metrics. For small-scale code, the F1 score increases by 5.5\% for Python and 6.1\% for Java, while overall accuracy rises by 6.2\% for Python and 8.0\% for Java. For larger-scale code, the verification and completion modules contribute an average F1 score improvement of 24.9\%. Additionally, \tool’s Accuracy-EM improves by 50\%, highlighting that generating program slices directly with a single agent often fails to produce correct slices. Iterative verification and refinement are therefore crucial for ensuring the accuracy of \tool.

\begin{figure}[!t]
	\centering
	\includegraphics[width=0.45\textwidth]{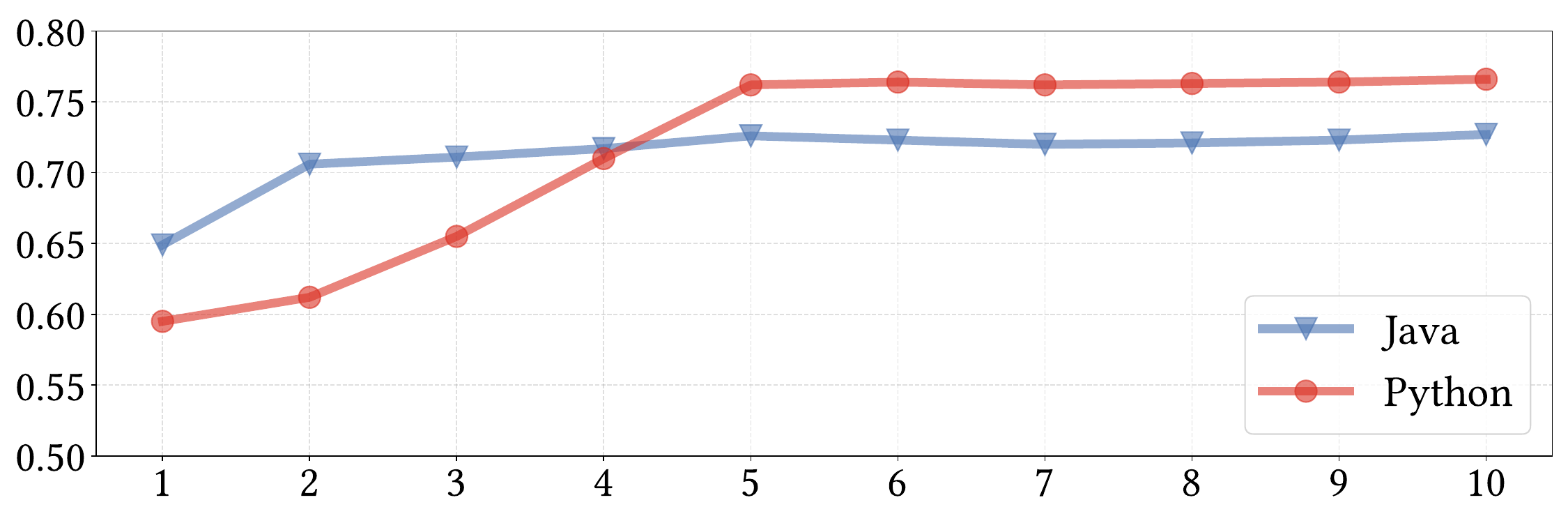}
	\vspace{-0.4cm}
	\caption{F1 scores of \tool for different control module iterations (RQ3). }
	\label{maxUp}
	\vspace{-0.4cm}
\end{figure}

\vspace{-0.1cm}
\subsubsection{Setting for Control Module}\label{subsec:Control} The control module uses a maximum iteration parameter to limit the execution of the verification and refinement agents. To determine an appropriate upper bound, we analyze how varying this maximum affects \tool's performance by testing iteration counts from 1 to 10. The results, shown in Figure \ref{maxUp}, display performance for Python (red line) and Java (blue line). We observe a steady increase in the F1 score as the iteration count grows, from 0.595 to 0.766. The most significant improvement occurs between 0 and 5 iterations, with gains of 0.167 for Python and 0.077 for Java. In contrast, the improvement from 5 to 10 iterations is marginal, with only an additional 0.04 for Python and 0.001 for Java. These results suggest that while the verification and refinement agents are effective, further iterations provide diminishing returns. Based on this trend, we set the default maximum iterations for \tool to 5.

\vspace{-0.1cm}
\begin{tcolorbox}[
	colback=gray!10, 
	colframe=black!70, 
	coltitle=black, 
	boxrule=0.75pt, 
	rounded corners, 
	drop shadow, 
	enhanced, 
	shadow={1mm}{-1mm}{0mm}{black!50}, 
	boxsep=0.1mm 
	]
	\textbf{Answer to RQ 3:} Both verification and refinement agents contribute to \tool’s performance. Notably, removing them results in a 5.5\%–29.1\% reduction in F1 score. Furthermore, setting the control module’s maximum number of iterations to 5 provides an optimal configuration.
\end{tcolorbox}
\vspace{-0.3cm}

\section{Threats to Validity}

\textbf{Internal Threats.}
For NS-Slicer, which was originally fine-tuned on slices produced by JavaSlicer.
Ideally, NS-Slicer should be further fine-tuned on our new benchmark, but the limited dataset size may affect its performance. To mitigate this, we applied cross-validation on the small-scale subset, allowing NS-Slicer to learn slicing patterns relevant to our benchmark. For large-scale programs, we selected the fine-tuned model with the highest F1 score to generate slices involving cross-file dependencies, preserving its inter-procedural capabilities as much as possible under realistic constraints. For buggy statement annotation, inconsistency in labeling criteria poses a potential threat. We mitigate it by following a widely adopted heuristic from prior studies~\cite{Meng2022, Pearson2017}, labeling deleted or modified lines in the fixing commit as buggy.

\vspace{0.1cm}
\noindent\textbf{External Threats.}
External threats concern the generalizability of our results. While \tool has been evaluated on Java and Python, it has not yet been applied to other popular languages such as C/C++, which may limit its applicability. Nevertheless, \tool is language-agnostic and can be adapted to other languages with appropriate front-end support. Another concern lies in benchmark coverage: although our dataset includes both small-scale examples and complex real-world projects, it cannot exhaustively capture the full diversity of coding styles and language constructs. To address this, we plan to continuously expand and refine the benchmark to improve its representativeness and coverage.

\section{Related Work}

\subsection{Traditional Static Program Slicing}

Traditional static program slicing techniques are generally divided into two categories: (1) those based on intermediate program representations and (2) those based on limited dependency traversal. In the first category, Ottenstein et al.~\cite{Ferrante1987} introduced the PDG, which models intra-procedural data and control dependencies, reducing slicing to a reachability problem in a graph. Horwitz et al.~\cite{Horwitz1990} extended this to inter-procedural slicing with the System Dependency Graph (SDG), which served as the foundation for subsequent techniques. For example, the hierarchical dependency graph~\cite{Li2004} encompasses package, class, function, and statement-level structures; the multi-threaded dependency graph~\cite{Zhao1999} includes inter-thread dependencies; and the sub-statement level system dependency graph~\cite{wang2020} considers the polymorphic features of Java.

In the second category, techniques enhance efficiency and scalability by limiting dependency traversal rather than constructing comprehensive representations. Krinke~\cite{Krinke2003} proposed barrier slicing, which halts dependency propagation at user-defined barriers in the graph. Sridharan et al.~\cite{Sridharan2007} introduced thin slicing, focusing on variables directly related to the slicing criterion. Palepu et al.~\cite{Palepu2014} assigned weights to graph nodes and edges to guide traversal. Wu et al.~\cite{Wu2024} further reduced complexity by selecting key variables and restricting analysis to strictly relevant ones, avoiding transitive dependencies. While the two types of slicing methods are effective, they require complete syntactic structures and are memory-intensive, limiting their practicality for large-scale programs. In contrast, \tool operates directly on raw source code without explicit intermediate representations, providing a lightweight and scalable alternative.

\vspace{-0.1cm}
\subsection{Learning-Based Static Program Slicing}
In recent years, the growing adoption of deep learning in software engineering has spurred interest in using neural models to enhance program slicing. Yadavally et al.~\cite{Yadavally2023} proposed NeuralPDA, which formulates dependency prediction as a binary classification task to detect data and control dependencies between statements, enabling the construction of PDG through deep learning. To further support inter-procedural and fine-grained slicing, Yadavally et al.~\cite{Yadavally2024} fine-tunes GraphCodeBERT to classify whether each statement belongs to a slice, learning slicing behavior from labeled data. Shahandashti et al.~\cite{Kimya2024} applied recent advanced LLMs (e.g., GPT‑4o) to program slicing and empirically found that simple prompting yields suboptimal results, which necessitates our work. 

Despite these advancements, existing LLM-based slicing methods still depend heavily on traditional slicing tools to construct training datasets due to the high cost of manual slice annotation. As a result, their performance is inherently limited by the quality and coverage of these tools. Additionally, most methods treat slicing as a statement-level binary classification task, which is computationally inefficient for large programs. \tool eliminates the need for labeled datasets by utilizing LLMs with extensive programming knowledge to analyze and generate program slices, enhancing scalability while maintaining high precision, even on incomplete code or large multi-file programs.

\vspace{-0.05cm}
\section{Conclusion and Future Work}
We present \tool, the first LLM-based multi-agent framework for static program slicing. \tool consists of three core agents—synthesis, verification, and completion—along with a control module that coordinates their collaboration to produce high-quality slices. We also develop dedicated structural prompts and analysis interfaces that allow \tool to extract and reason over program information without manual intervention. For rigorous evaluation, we introduce \bench, a manually curated benchmark comprising 2,200 slices across Java and Python programs. We compare \tool against representative traditional tools and state-of-the-art learning-based methods. Experimental results show that \tool consistently outperforms all baselines, achieving average improvements of 15.2\% in Accuracy, 22.1\% in F1 score, and 10.7\% in Accuracy-EM on \bench.

While this work focuses on static slicing, future efforts will explore extending \tool to support dynamic slicing, enabling analysis of inter-run dependency propagation using runtime information. 
\newpage

\balance
\bibliographystyle{unsrt}
\bibliography{reference}

\end{document}